# Low-frequency electromagnetic radiation of hydrogen molecular gas: effect of the ortho-para conversion

A.M. Ishkhanyan[1,2], and V.P. Krainov[3]

[1]Russian-Armenian University, H. Emin 123, 0051, Yerevan, Armenia
[2]Institute of Physics and Technology, National Research Tomsk Polytechnic University, Tomsk, 634050 Russian Federation
[3]Moscow Institute of Physics and Technology, Dolgoprudny 141700, Russian Federation

e-mail: vpkrainov@gmail.com

**Abstract**. We calculate the rate of the spontaneous magnetic dipole radiation transitions from ortho- to para-states of the hydrogen molecule at the room temperature with the radiation wavelengths about $0.01 - 0.1$ cm. Exponentially small differences between the rotational energies and heat capacities of parahydrogen and orthohydrogen molecules and also diatomic molecules consisting of different hydrogen atoms for high temperatures are derived.

1. **Introduction**

The hydrogen molecule is one of the most fundamental molecules in chemistry, biology and astrophysics. It exists as two distinct nuclear-spin isomers, para- and orthohydrogen. But known experimental works [1-3] deal with water, not hydrogen. Authors of Ref. [4] report the formation of a magnetically focused beam of ortho-water. Authors of Ref. [5] report the results of an experimental study related to the relaxation of the nuclear spin isomers of the water molecule in a supersonic expansion. Methods to separate hydrogen molecules into para- and ortho-isomers have allowed important insights into the different physical properties of these two spin isomers [6]. In the work [7] authors study the possibility of observing strongly forbidden vibrational–rotational transitions between ortho and para isotopomers for water. These are yet to be observed experimentally: From such an observation it would be possible to develop reliable models about the ortho and para interconversion mechanism. This is important, for example, in the study of astrophysical cometary problems. In the dense hydrogen gas collisions between molecules lead to ortho-para conversion in hydrogen by collisions with the oxygen molecule [8]. We calculate the rate of the spontaneous magnetic dipole radiation transitions from ortho- to para-states of the dilute hydrogen gas at the room temperature. It is interesting since the energy difference between these states cannot be derived by standard asymptotical methods. This difference is exponentially small while the known asymptotic power expressions coincide for ortho- and para-states. In order to derive this difference, we use fine approach based on the application of the Poisson formula

$$\sum_{n=0}^{\infty} F(n) = \int_{0}^{\infty} F(x)dx + \frac{1}{2}F(0) + 2\sum_{n=1}^{\infty} \int_{0}^{\infty} F(x)\cos(2\pi nx)dx.$$

**2. Heat capacities**

Rotational partition functions for molecules of orthohydrogen ($o$) and parahydrogen ($n$) are given by known expressions (with respect to one molecule) [10]

$$Z_o = \sum_{n=0}^{\infty}(4n+3)\exp\left[-(n+1)(2n+1)/x\right], \qquad (1)$$

$$Z_n = \sum_{n=0}^{\infty}(4n+1)\exp\left[-n(2n+1)/x\right], \qquad (2)$$

where $x = IT/\hbar^2$, $I$ is the molecular moment of inertia, and $T$ is temperature. In Eq. (1) we do not take into account the factor 3 due to the spin degeneration, since it is inessential at the derivation of thermodynamic energies and heat capacities. It is known that in both cases the expressions lead to the same asymptotic expansions for the rotational partition functions, energies and for the heat capacities at high temperatures [10]:

$$C_n = C_o = 1 + \frac{1}{45}\left(\frac{T_0}{T}\right)^2 + \frac{16}{945}\left(\frac{T_0}{T}\right)^3 + \ldots, \qquad (3)$$

where $T_0 = \hbar^2/2I$ is the characteristic rotational molecular energy. Therefore, it is of interest to find the difference between these heat capacities at high temperatures $T \gg T_0$.

To examine the difference between the rotational partition functions, we first separate the most rapidly varying factor $\exp(-2n^2/x)$:

$$Z_n - Z_o = \sum_{n=0}^{\infty} g(n,x)\exp\left(-\frac{2n^2}{x}\right), \qquad (4)$$

where $g(n,x)$ is a smoother function compared with this factor. In order to estimate this difference we use the known Poisson formula [10] and restrict ourselves to the first harmonic only (the other harmonics lead to much less contributions)

$$Z_n - Z_o \approx \mathrm{Re}\left(\int_{-\infty}^{\infty} g(n,x)\exp\left(-\frac{2n^2}{x} + 2i\pi n\right)dn\right). \qquad (5)$$

When $x \gg 1$, this integral can be evaluated by the saddle-point method. The notation is introduced

$$f(n) = -\frac{2n^2}{x} + 2i\pi n. \qquad (6)$$

Then
$$f'(n) = -\frac{4n_0}{x} + 2i\pi = 0 \Rightarrow n_0 = \frac{i\pi x}{2}, \tag{7}$$

so that
$$f(n_0) = -\frac{\pi^2 x}{2}, \quad f''(n_0) = -4/x \tag{8}$$

and
$$\sqrt{\frac{2\pi}{|f''(n_0)|}} = \sqrt{\frac{\pi x}{2}}. \tag{9}$$

Hence, the saddle-point result is
$$Z_n - Z_o \approx \text{Re}\left(g(n_0)\sqrt{\pi x/2} \cdot e^{f(n_0)}\right). \tag{10}$$

Using equations (1), (2), for $g(n_0)$ we have
$$g(n_0) = (4n_0 + 1)\exp\left(-\frac{n_0}{x}\right) - (4n_0 + 3)\exp\left(-\frac{3n_0}{x} - \frac{1}{x}\right). \tag{11}$$

With $n_0 = i\pi x/2$, we arrive at $x \gg 1$:
$$g(n_0) = -i(2\pi i x + 1) - i(2\pi i x + 3)\exp\left(-\frac{1}{x}\right) \approx 4\pi x. \tag{12}$$

Thus, the pre-exponential factor in equation (10) is equal to $(2\pi x)^{3/2}$, and the overall result reads
$$Z_n - Z_o \approx (2\pi x)^{3/2} \exp\left(-\frac{\pi^2 x}{2}\right). \tag{13}$$

In the zero-order approximation $Z_o = x$. Therefore, we obtain
$$Z_n \approx Z_o + (2\pi x)^{3/2} \exp\left(-\frac{\pi^2 x}{2}\right) \approx Z_o\left(1 + (2\pi)^{3/2} x^{1/2} \exp\left(-\frac{\pi^2 x}{2}\right)\right). \tag{14}$$

Since $x \gg 1$, then
$$(2\pi)^{3/2} x^{1/2} \exp\left(-\frac{\pi^2 x}{2}\right) \ll 1.$$

Hence, using the expansion $\ln(1+\alpha) \approx \alpha$ at $\alpha \ll 1$, one obtains
$$\ln Z_n \approx \ln Z_o + (2\pi)^{3/2} x^{1/2} \exp\left(-\frac{\pi^2 x}{2}\right). \tag{15}$$

This gives the exponentially small difference of the Helmholtz free energies as
$$\Delta F_{no} = -T(\ln Z_n - \ln Z_o) = -(2\pi)^{3/2} T x^{1/2} \exp\left(-\frac{\pi^2 x}{2}\right). \tag{16}$$

Here, one may differentiate only the exponential factor, since this will lead to the principal contribution. Now we find the difference of energies of para- and ortho-hydrogen molecules:

$$E = F - T\frac{dF}{dT}, \qquad (17)$$

$$\Delta E_{on} = \pi^{7/2} x^{3/2} \sqrt{2} T \exp\left(-\frac{\pi^2 x}{2}\right). \qquad (18)$$

At the room temperature we have $I = mR^2/2$; here $m$ is the proton mass, the equilibrium distance between the protons is $R = 0.74$ Å, $x = 1.70$, and $\Delta E_{on} = 0.040T = 3.8 \cdot 10^{-5}$ a.u. $\approx 1 \cdot 10^{-3}$ eV. The corresponding wavelength of the spontaneous radiation is $\lambdabar = 2\pi c\hbar / \Delta E_{on} = 2.26 \cdot 10^7$ a.u. $= 0.12$ cm. According to experimental data we have in the statistical equilibrium at the room temperature the chemical reaction [9]

$$o\text{H}_2 \rightarrow n\text{H}_2 + 340 \text{ cal/mol} .$$

Hence, $\Delta E_{on}^{\exp} = 340 \text{ cal/mol} = 1.47 \cdot 10^{-2}$ eV and the wavelength is $\lambdabar = 80$ μm. It is seen that the experimental value of the energy difference is much more than the calculated difference of the averaged energies, Eq. (18). The reason is that the quantity $x = 1.70$ for room temperature is not too high to apply asymptotic expansions which depend exponentially on this dimensionless parameter. Exact theoretical value of $\Delta E_{on}$ is

$$\Delta E_{on}^{theor} = 2T_0 x^2 \frac{d}{dx}\left(\ln \frac{Z_o}{Z_n}\right).$$

Substituting Eqs. (1) and (2) for rotational partition functions, we can derive numerically this difference for the value of $x = 1.70$, and the result coincides the experimental value cited above, as it should be.

Thus, the small difference of heat capacities is:

$$\Delta C_{no} \approx \frac{\pi^{11/2}}{\sqrt{2}} x^{5/2} \exp\left(-\frac{\pi^2 x}{2}\right). \qquad (19)$$

In particular, for $x = 7$ we find $\Delta C_{no} = 4.9 \cdot 10^{-11}$. The exact numerical evolution gives a very close value: $\Delta C_{no} = 4.5 \cdot 10^{-11}$. Of course, from practical point of view we should have in mind the strong dissociation of hydrogen molecules at these high temperatures.

It follows from Eq. (19) that

$$\frac{d}{dx}\ln \Delta C_{no} \approx -\frac{\pi^2}{2} + \frac{5}{2x}. \qquad (20)$$

In Fig. 1 we compare the dependence given by this equation with the numerical result as a function of $x \gg 1$. One can see a good agreement. Thus, the exponentially small difference between the heat capacities of parahydrogen and orthohydrogen molecules is approximately given by the simple analytic expression (19).

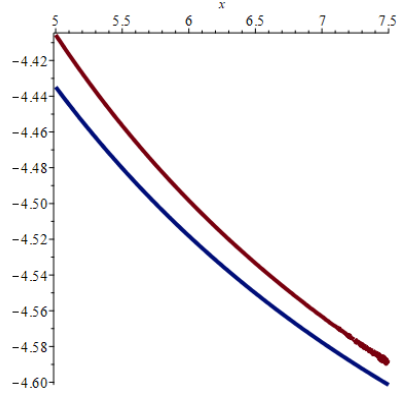

Fig. 1. Comparison of the analytic dependence for the logarithmic derivative of the heat capacity difference, equation (20), (lower curve) with the exact numerical result (upper curve).

We now determine the difference of the rotational heat capacities between a diatomic molecule consisting of different atoms and the parahydrogen molecule at high temperatures. The asymptotic expansion of the heat capacity for a molecule consisting of different atoms is also given by equation (3) [11]. The rotational partition function for a diatomic molecule is $Z_r = Z_n + Z_0$. Hence, its logarithm is equal to

$$\ln Z_r = \ln Z_n + \ln\left(1 + \frac{Z_o}{Z_n}\right) = \ln Z_n + \ln\left[2\left(1 + \frac{Z_o - Z_n}{2Z_n}\right)\right].$$

Since $(Z_o - Z_n)/Z_n \ll 1$, one obtains

$$\ln Z_r \approx \ln Z_n + \ln 2 + \frac{Z_o - Z_n}{2Z_n} \approx \ln Z_n + \ln 2 + \frac{Z_o - Z_n}{2x}. \tag{21}$$

We have used here that according to Eq. (10)

$$Z_n \approx 4\int_0^\infty n\,dn \cdot \exp(-2n^2/x) = x.$$

For the difference of heat capacities between the molecule consisting of different atoms $C_r$ and the parahydrogen molecule we then have

$$\Delta C_{rn} = x\frac{d^2}{dx^2}\left(x\left(\ln Z_r - \ln Z_n\right)\right) = -\frac{x}{2}\frac{d^2}{dx^2}(Z_n - Z_0). \tag{22}$$

With equation (13), we obtain

$$\Delta C_{rn} \approx -\frac{(2\pi)^{3/2} x}{2} \frac{d^2}{dx^2}\left( x^{3/2} \exp\left(-\frac{\pi^2 x}{2}\right)\right). \tag{23}$$

Here again one may differentiate only the exponential factor, since this will lead to the principal contribution. The final result is

$$\Delta C_{rn} = -\frac{\pi^{11/2} x^{5/2}}{2\sqrt{2}} \exp\left(-\frac{\pi^2 x}{2}\right). \tag{24}$$

It follows from equations (19) and (24) that the difference between heat capacities of the molecule consisting of different atoms, and the orthohydrogen molecule is

$$\Delta C_{ro} = \Delta C_{rn} + \Delta C_{no} = \frac{\pi^{11/2} x^{5/2}}{2\sqrt{2}} \exp\left(-\frac{\pi^2 x}{2}\right). \tag{25}$$

Thus, at a given high temperature the value of the heat capacity for the molecule consisting of two different atoms is right in the middle of the segment between the heat capacities of the ortohydrogen and parahydrogen.

3. **Spontaneous radiation between the ortho- and para-states**

Now we consider the spontaneous radiation transitions from the ortho-state to the lower lying para-state of the hydrogen molecule. The vector potential **A** of the vacuum field is of the form ($v$ is the normalization volume):

$$\mathbf{A}(\mathbf{r},t) = c\mathbf{e}\sqrt{\frac{2\pi\hbar}{v\omega}} \hat{a}^+ \exp(-i\mathbf{k}\mathbf{r} + i\omega t). \tag{26}$$

Here **e** is the unit polarization vector of the emitted photon. In the dipole approximation, we have $\mathbf{kr} \ll 1$. The second term of the Taylor expansion of this equation is

$$\mathbf{A}(\mathbf{r},t) = -c\mathbf{e}(\mathbf{kr})^2 \sqrt{\frac{\pi\hbar}{2v\omega}} \exp(i\omega t). \tag{27}$$

The first term of the Taylor expansion does not contribute to the rate of the spontaneous emission because of the orthogonality of the spatial wave functions of para- and ortho-states, since then the magnetic field strength **H** does not depend on the proton coordinates. The matrix element of the operator of the photon birth is $\langle 1|\hat{a}^+|0\rangle = 1$. Besides of this, $k = \omega/c$.

Now we determine the magnetic field strength of the vacuum field:

$$\mathbf{H} = \text{rot}\mathbf{A} = -x\cos\theta[\mathbf{k},\mathbf{e}]\sqrt{2\pi\hbar\omega/v} \exp(i\omega t). \tag{28}$$

Here the angle $\theta$ determines the direction of the emitted photon with respect to the molecular axis $X$. Then $k_x = k\cos\theta$, $k_y = k\sin\theta$, $k_z = 0$. The photon wave vector **k** is normal to the polarization vector **e**. Then $e_x = -\sin\theta\cos\varphi$, $e_y = \cos\theta\cos\varphi$, $e_z = \sin\varphi$ and

$$[\mathbf{k},\mathbf{e}]_z = k\cos\varphi, \quad [\mathbf{k},\mathbf{e}]_x = k\sin\theta\sin\varphi, \quad [\mathbf{k},\mathbf{e}]_y = -k\cos\theta\sin\varphi. \tag{29}$$

The interaction of the magnetic moment of the hydrogen molecule with the magnetic vacuum field is given by

$$\hat{V} = -\hat{\mathbf{M}}\mathbf{H} = -\mu_B g\left(\hat{\boldsymbol{\sigma}}_1 \mathbf{H}(\mathbf{r}_1) + \hat{\boldsymbol{\sigma}}_2 \mathbf{H}(\mathbf{r}_2)\right), \tag{30}$$

where $\mu_B = e\hbar/2mc$ is the Bohr proton magneton, $m$ is the proton mass, $g$ is the proton gyromagnetic factor, and $\hat{\boldsymbol{\sigma}}_1, \hat{\boldsymbol{\sigma}}_2$ are the Pauli matrices for the first and the second protons in the hydrogen molecule, respectively. We have

$$\hat{\boldsymbol{\sigma}}_1 [\mathbf{k},\mathbf{e}] = k\begin{pmatrix} \cos\varphi & \sin\varphi\exp(i\theta) \\ \sin\varphi\exp(-i\theta) & -\cos\varphi \end{pmatrix}_1. \tag{31}$$

The orthogonal spin functions of the para- and ortho-states are:

$$\psi_n^{spin} = \frac{1}{\sqrt{2}}\left\{\begin{pmatrix}1\\0\end{pmatrix}_1 \begin{pmatrix}0\\1\end{pmatrix}_2 + \begin{pmatrix}1\\0\end{pmatrix}_2 \begin{pmatrix}0\\1\end{pmatrix}_1\right\}, \tag{32}$$

$$\psi_o^{spin} = \frac{1}{\sqrt{2}}\left\{\begin{pmatrix}1\\0\end{pmatrix}_1 \begin{pmatrix}0\\1\end{pmatrix}_2 - \begin{pmatrix}1\\0\end{pmatrix}_2 \begin{pmatrix}0\\1\end{pmatrix}_1\right\}. \tag{33}$$

For the spin parts of the transition matrix element we then obtain

$$\left\langle \psi_o^{spin} \middle| \hat{\boldsymbol{\sigma}}_1 [\mathbf{k},\mathbf{e}] \middle| \psi_n^{spin} \right\rangle = -\left\langle \psi_o^{spin} \middle| \hat{\boldsymbol{\sigma}}_2 [\mathbf{k},\mathbf{e}] \middle| \psi_n^{spin} \right\rangle = k\cos\varphi. \tag{34}$$

Now we consider the spatial wave functions of protons, which oscillate around their equilibrium positions. We assume one-dimensional harmonic oscillations along the molecular axis with the frequency $\Omega$ and with small amplitudes compared to the equilibrium distance $2R$ between protons in the hydrogen molecule. The oscillator one-dimensional wave function of the ground state is (the $X$ axis is chosen along the molecular axis)

$$\psi(x,t) = \left(\frac{m\Omega}{\pi\hbar}\right)^{1/4} \exp\left(-\frac{m\Omega x^2}{2\hbar} - \frac{i\Omega t}{2}\right), \tag{35}$$

where $m$ is the proton mass. Hence, the spatial wave functions of para- and ortho-states which are orthogonal each to other, are

$$\psi_o(x_1, x_2, t) = \left(\frac{m\Omega}{2\pi\hbar}\right)^{1/2} \exp(-iE_o - i\Omega t) \times$$

$$\times \left\{\exp\left(-\frac{m\Omega\left((x_1+R)^2+(x_2-R)^2\right)}{2\hbar}\right) + \exp\left(-\frac{m\Omega\left((x_2+R)^2+(x_1-R)^2\right)}{2\hbar}\right)\right\}, \quad (36)$$

$$\psi_n(x_1, x_2, t) = \left(\frac{m\Omega}{2\pi\hbar}\right)^{1/2} \exp(-iE_n - i\Omega t) \times$$

$$\times \left\{\exp\left(-\frac{m\Omega\left((x_1+R)^2+(x_2-R)^2\right)}{2\hbar}\right) - \exp\left(-\frac{m\Omega\left((x_2+R)^2+(x_1-R)^2\right)}{2\hbar}\right)\right\}. \quad (37)$$

Here we took into account that $R \gg x_1, x_2$, due to smallness of the oscillation amplitudes. The energy difference $\Delta E_{on} = E_o - E_n$ is given by equation (18), or (better) by the experimental value.

We now derive the spatial dipole matrix elements:

$$\langle \psi_o(x_1,x_2,t)|x_1|\psi_n(x_1,x_2,t)\rangle = -\langle \psi_o(x_1,x_2,t)|x_2|\psi_n(x_1,x_2,t)\rangle =$$
$$= -2R\exp(-i\Delta E_{on}). \quad (38)$$

We note that the frequency $\Omega$ does not appear in this result. Thus, the transition matrix element is

$$\hat{V}_{on} = 2R\mu_B g\sqrt{\frac{2\pi\hbar\omega^3}{vc^2}}\cos\theta\cos\varphi\exp(-i\Delta E_{op}). \quad (39)$$

With this, the rate of the spontaneous magnetic dipole transition determined by the "golden Fermi rule" reads

$$w_{on} = \frac{2\pi}{\hbar}\int|\hat{V}_{on}|^2 \delta(\hbar\omega - \Delta E_{on})\frac{2vd\mathbf{k}}{(2\pi)^3} = \frac{4\pi e^2\hbar g^2}{3m^2c^2R^3}\left(\frac{R\Delta E_{on}^{\exp}}{\hbar c}\right)^5, \quad (40)$$

It follows from Eq. (40) at the room temperature that $1/w_{on} = 2.5 \cdot 10^{35}$ c $\sim 10^{25}$ years. Hence, for observation of the spontaneous radiation we should have more than about $10^{23}$ hydrogen molecules in the Earth atmosphere.

**Conclusion**

The rate of the spontaneous magnetic dipole transition from molecular orthohydrogen to molecular parahydrogen is calculated. Spin flip and dipole transition of the spatial part of the molecular wave function lead to nonzero transition matrix element. The transition rate does not depend on the frequency of the small proton oscillations around their equilibrium positions. We obtained also simple analytic expression for the exponentially small difference in thermodynamic energies and heat capacities between orthohydrogen and parahydrogen at high temperatures. Asymptotic limit is achieved for temperatures that are higher compared to the room temperature.

**Acknowledgements**

The work has been supported by the Armenian State Committee of Science (SCS Grant No. 18RF-139), the Armenian National Science and Education Fund (ANSEF Grant No. PS-4986), the Russian-Armenian (Slavonic) University at the expense of the Ministry of Education and Science of Russian Federation, the project "Leading Russian Research Universities" (Grant No. FTI_24_2016 of the Tomsk Polytechnic University), the Russian Foundation for Basic Research (project № 18-52-05006), and the Ministry of Education and Science of Russian Federation (project № 3.873.2017/4.6).

**References**

1. Kilaj A et al 2018 *Nature Communications* **9** 2096
2. Horke D A et al 2014 *Angew. Chem. Int. Ed*. **53** 11965
3. Tikhonov V I and Volkov A A 2002 *Science* **296** 2363
4. Kanamuri H et al 2017 *Phys. Rev. Lett.* **119** 173401
5. Georges R et al 2017 *J. Phys. Chem*. A **121** 7455
6. Kravchuk T et al 2011 *Science* **331** 319
7. Cacciani P, Cosleou J and Khelkhal M 2012 *Phys. Rev.* A **85** 012521
8. Mineev B F and Agren H 1995 *J. Phys. Chem.* **99** 8936
9. Miani A and Tennyson J 2004 *J. Chem. Phys.* **120** 2732
10. Kubo R *Statistical Mechanics*, North-Holland Publishing Company, Amsterdam, 7$^{th}$ edition, 1988.
11. Landau L D and Lifshitz E M *Statistical Physics*, Volume 5 (Course of Theoretical Physics), Elsevier, Amsterdam, 3d edition, 1980.